# Unusual thickness dependence of exciton characteristics in 2D perovskite quantum wells

J.-C. Blancon[1]*, A. V. Stier[1], H. Tsai[1,2], W. Nie[1], C. C. Stoumpos[3], B. Traoré[4], L. Pedesseau[5], M. Kepenekian[4], S. Tretiak[1], S. A. Crooker[1], C. Katan[4], M. G. Kanatzidis[3,6], J. J. Crochet[1], J. Even[5]* and A. D. Mohite[1]*

[1]Los Alamos National Laboratory, Los Alamos, New Mexico 87545, USA.

[2]Department of Materials Science and Nanoengineering, Rice University, Houston, Texas 77005, USA.

[3]Department of Chemistry, Northwestern University, Evanston, Illinois 60208, USA.

[4]Institut des Sciences Chimiques de Rennes (ISCR), Université de Rennes 1, CNRS, Ecole Nationale Supérieure de Chimie de Rennes, INSA de Rennes, UMR6226, 35042 Rennes, France.

[5]Fonctions Optiques pour les Technologies de l'Information (FOTON), INSA de Rennes, CNRS, UMR 6082, 35708 Rennes, France.

[6]Department of Materials Science and Engineering, Northwestern University, Evanston, Illinois 60208, USA.

*Correspondence to: jblancon@lanl.gov, jacky.even@insa-rennes.fr, amohite@lanl.gov

**Understanding the nature and energy distribution of optical resonances is of central importance in low-dimensional materials and its knowledge is critical for designing efficient optoelectronic devices. Ruddlesden-Popper halide perovskites are 2D solution-processed quantum wells with a general formula $A_2A'_{n-1}M_nX_{3n+1}$, where optoelectronic properties can be tuned by varying the perovskite layer thickness (n value), and have recently emerged as efficient semiconductors with technologically relevant stability. However, fundamental questions concerning the nature of optical resonances (excitons or free-carriers) and the exciton reduced mass, and their scaling with quantum well thickness remains unresolved. Here, using optical spectroscopy and 60-Tesla magneto-absorption supported by modelling, we unambiguously demonstrate that the optical resonances arise from tightly bound excitons with unexpectly high exciton reduced mass (0.20 $m_0$) and binding energies varying from 470 meV to 125 meV with increasing thickness from n=1 to 5. Our work demonstrates the dominant role of Coulomb interactions in 2D solution-processed quantum wells and presents unique opportunities for next-generation optoelectronic and photonic devices.**

Ruddlesden-Popper halide perovskites[1,2] (RPPs) are solution-processed quantum well structures formed by two-dimensional (2D) layers of halide perovskite semiconductors separated by bulky organic spacer layers, whose stoichiometric ratios are defined by the general formula[3] $A_2A'_{n-1}M_nX_{3n+1}$ where A, A' are cations, M is a metal, X is a halide, and the integer value n determines the perovskite layer thickness (or quantum well thickness). Recent breakthrough in the synthesis of phase-pure (a single n-value) RPPs with higher values[3–5] of n, up to n=5, has inspired their use as low-cost semiconductors in optoelectronics[5–8] as an alternative to three-dimensional (3D) perovskites due to their technologically relevant intrinsic photo- and chemical-stability[5–10]. However, key fundamental questions remain unanswered in RPPs with n>1, such as the nature of optical transitions, as well as the behaviour of Coulomb interactions especially with increasing quantum well thickness. In fact there has been an intense ongoing debate[6–8,11–13] regarding the exact nature of the optical transitions (excitons versus free carriers) in RPPs with large n-values. RPPs with n=1,[14,15] (excitons at RT) and 3D perovskites[16] (free carriers at RT) are representative of the two limiting regimes at room temperature, but the analysis of the crossover has not been performed. This issue originates from the lack of knowledge of the fundamental quantities such as the exciton reduced mass, dielectric constant and characteristics like the spatial extend of electron and hole wavefunctions, which play a crucial role in the determination of the exciton binding energy. In particular, contradictory reports of the value of the exciton reduced mass have significantly contributed to this uncertainty[15,16] making it the most critical experimentally derived parameter required for the quantitative determination of the exciton characteristics in hybrid (organic-inorganic) perovskite systems. Moreover, unlike bulk inorganic semiconductors, heterostructures[17] and 3D perovskites[18,19], it is non-trivial to determine the exciton reduced mass in layered hybrid perovskites using a symmetry based (k.p) approach[20] or using many-body ab-initio calculations[21]. Furthermore, ambiguity in the determination of the exciton binding energy in layered 2D perovskites has also been imposed by the limited understanding of the role of dielectric confinement versus quantum confinement with increasing quantum well thickness[22,23]. More generally, phase-pure RPPs with n>1 present a unique opportunity to explore the physical properties of natural quantum-well semiconducting crystals intermediate between monolayer 2D materials[24,25] and 3D materials[26], a quasi-dimensional physics only accessible in synthetic inorganic semiconductor quantum-well structures so far[26].

Here we present the first study on using low-temperature (4K) magneto-optical spectroscopy to accurately determine the exciton reduced mass for $(BA)_2(MA)_{n-1}Pb_nI_{3n+1}$ RPP crystals with perovskite layer thickness varying between n=1 and n=5 (Fig. 1a, Supplementary Fig. 1, and Supplementary Table 1). The reduced mass is used to develop a generalized theoretical model for electron-hole interactions in RPPs and determine fundamental characteristics of the exciton states. In parallel, from low temperature optical spectroscopy we experimentally determine the exciton binding energy in the RPPs n=1 to 5. Finally, from these results we produce a general scaling behaviour for the binding energy of Wannier-Mott exciton states in RPPs, which allows for prediction of the exciton binding energy for any given thickness. This study closes a long-standing scientific gap and will lead to the rational design of next-generation layered 2D perovskite based optoelectronic devices.

## Results

**Magneto-absorption measurement to determine the exciton reduced mass.** Fig. 1 shows the RPP structure and results from magneto-absorption spectroscopy, which was employed to probe the strength of the electron-hole interaction in RPPs and deduce the exciton reduced mass. These measurements were performed in the Faraday geometry at 4 K, and the optical spectra of the right-and left-handed circular polarization components ($\sigma^{\pm}$) of the transmitted white light were probed. Application of a high magnetic field on the RPP n=4 results in an energy shift of the optical resonance at ~1.9 eV (Fig. 1b, c). Similar results were obtained for the other RPPs (see Supplementary Fig. 2). This optical transition was identified as the exciton ground state and its energy shift under magnetic field in the Faraday configuration is expressed by[26] $\Delta E=1/2\ g_0\mu_B B + c_0B^2$, where the first term describes the Zeeman splitting of the $\sigma^+$ and $\sigma^-$ exciton transitions ($g_0$ is the g-factor in the perovskite plane, $\mu_B$ the Bohr magneton, B the magnetic field) and the second one the diamagnetic shift ($c_0$ is the diamagnetic shift coefficient). Fitting the exciton energy shifts for both $\sigma^+$ and $\sigma^-$ polarizations using the model above yields the diamagnetic coefficients $c_0$ of the RPPs with n>1 (Fig. d). Here, $c_0$ increases monotonically with the perovskite layer thickness, and it ranges from ~0.4 μeV/T$^2$ for the RPP n=2 to ~1.1 μeV/T$^2$ for n=5. The value for n=1 was obtained from Tanaka et al.[15] ($c_0$=0.26 μeV/T$^2$). On the other hand, the g-factor presents little dependence on the n value, with average $g_0$=1.5±0.1, which is consistent with reports on 3D perovskites[27,28]. Under magnetic field the competition between the Zeeman and diamagnetic effects explains the non-symmetric, opposite-sign energy shift of the $\sigma^+$ and $\sigma^-$ exciton transitions with respect to the zero field exciton absorption energy (Fig. 1c and Supplementary Fig. 2). We note that this asymmetry becomes more pronounced for thicker perovskite layers (n→5), which is due to the strong increase of the diamagnetic coefficient with increasing n value.

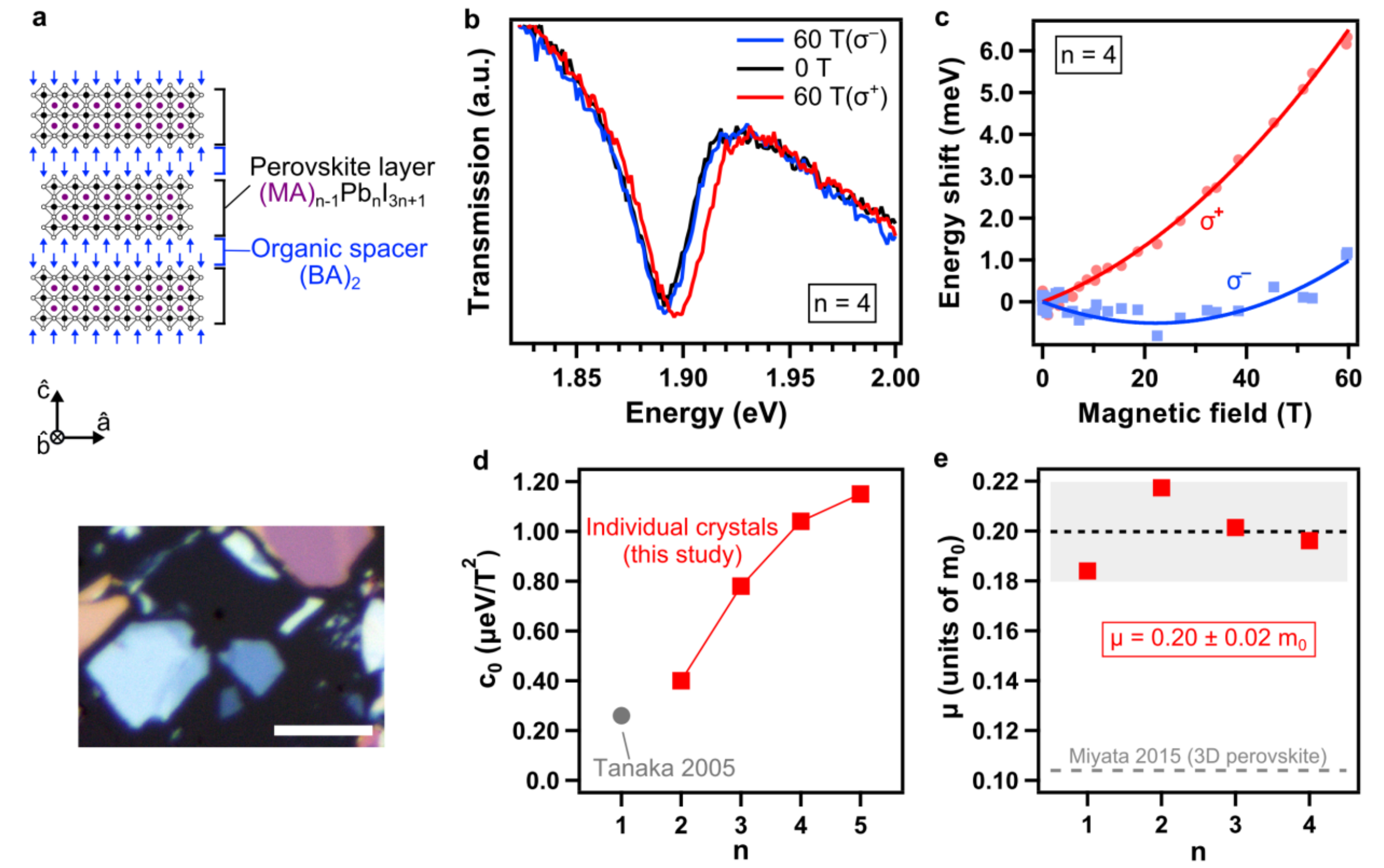

**Fig. 1. Exciton reduced mass from magneto-absorption spectroscopy and theory. a,** (top) Schematic of the RPP structure cut along the direction **ĉ** of stacking of the 2D layers. (bottom) Image of mechanically exfoliated RPP crystals. Scale bar is 10 µm. **b,** Magnetic field dependence of the light transmission of an individual RPP n=4 crystal for right- ($\sigma^-$) and left-handed ($\sigma^+$) circular polarization. **c,** Corresponding shift of the exciton energy as a function of the magnetic field. Fit of the data using $\Delta E=1/2\ g_0\mu_B B + c_0B^2$ yields $c_0$=1.04±0.16 µeV/T$^2$ and $g_0$=1.59±0.03. **d,** Derived diamagnetic shift coefficient of the measured RPPs (red squares). The value from Tanaka et al.[15] was used for the RPP n=1. **e,** Exciton reduced mass derived from fitting the experimental diamagnetic shifts with our theoretical model. The average value for RPPs is $\mu$=0.20±0.02 $m_0$. (Gray dashed line) 3D perovskite value measured by Miyata et al.[16]. Theory does not show n=5 due to computational limitations.

The diamagnetic coefficient is directly connected to both the exciton reduced mass and the strength of the electron-hole Coulomb interaction. Although a simple relation exists for pure 2D systems where quantum confinement dominates[22,26], this model does not apply to the RPPs due to the following reasons. First, the perovskite layer thickness is comparable to the spatial extent of the excitons[15,16,29] and the exciton wavefunction cannot be strictly confined to a 2D plane (see also Supplementary Text 1). Second, the dielectric confinement plays a key role in the photo-physics of RPPs[15,23,29]. Therefore, we developed a theoretical model (see details in the next section), which describes the electron-hole Coulomb interaction in thin semiconductors and includes dielectric confinement (Keldysh theory[22]). For this theoretical model to work, it requires an accurate determination of the exciton reduced mass (labelled μ). We evaluated the reduced mass for each RPP by adjusting the theoretical values of $c_0$ to those measured experimentally (Fig. 1e and Supplementary Text 1). This yielded μ=0.18 $m_0$ ($m_0$ is the free electron mass) for the RPP n=1, up to ~0.22 $m_0$ for n=2, and about 0.20 $m_0$ for n=3 and n=4. Due to computational limitations, our model was not applied to the RPP with n=5 for which the average value $\mu$=0.20 $m_0$ was used (Fig. 1e). The deduced values of exciton reduced mass are about twice as large as those reported for 3D perovskites[16], even for the RPPs with n=5 in which case the exciton states are well confined within the perovskite layer (see next section). This is surprising because a decrease of the exciton reduced mass would be expected for increasing perovskite layer thickness ($n \rightarrow 5$), with limiting value of 0.104 $m_0$ (3D perovskite), as the vicinity of the exciton approaches that of the 3D perovskites. Therefore, the unambiguous determination of the reduced mass allows us to develop a predictive model for excitons in solution-processed quantum well systems and calculate the exciton binding energies of the RPP compounds.

**Calculating the exciton binding energy of Wannier-Mott exciton in 2D perovskites.** The binding energy of the exciton ground state was calculated for each RPP (n=1 to 4) by inputting each respective value of the exciton reduced mass into our theoretical model and semi-empirically solving the Bethe-Salpeter equation based on the effective mass Green's function approach[23]. This approach includes: *(i)* calculating the electronic structures of the RPPs to extract exciton wavefunctions and dielectric confinement profiles using density functional theory (DFT), *(ii)* generalizing the Keldysh theory for 2D perovskites, and *(iii)* combining both of the above by building a semi-empirical model to simulate the Wannier-Mott exciton characteristics (Supplementary Text 1). In our model, the potential function describing the Coulomb interaction between the electron and hole forming the exciton states is based on Keldysh theory[22] generalized to a semiconducting dielectric quantum well (Fig. 2a), *i.e.* a dielectric well (perovskite layer) with thickness $d$ and dielectric constant $\varepsilon_w$ sandwiched between well barriers (organic spacing layers) with dielectric constant $\varepsilon_b$. The dielectric constants were taken at optical frequency because the values of binding energy measured in RPPs (see next section) are one order of magnitude larger than the highest energy value reported for the lattice optical phonon modes[30].

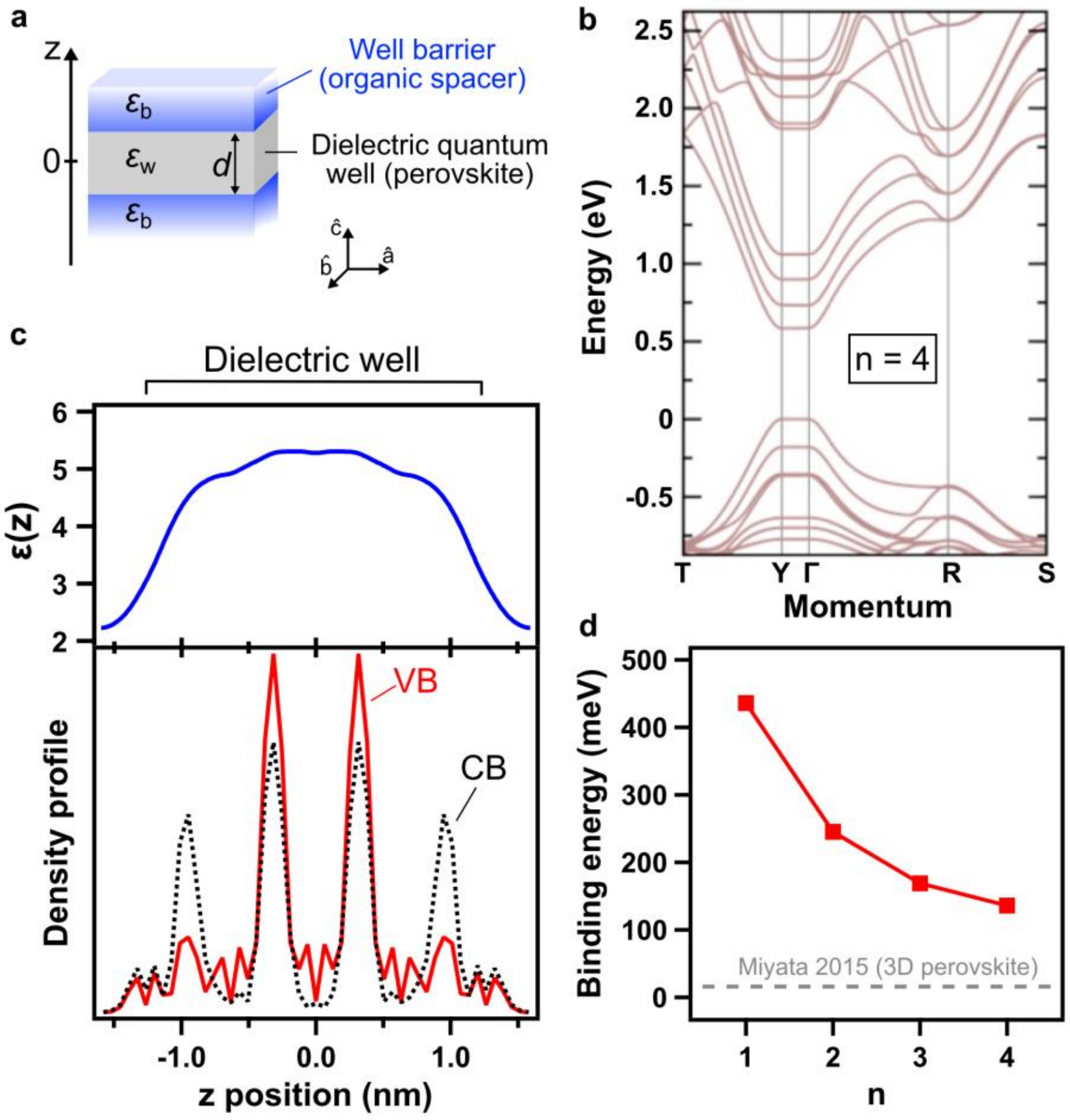

**Fig. 2 Semi-empirical model of Wannier-Mott exciton in RPPs. a,** Schematics of the single quantum well system to which our model was applied to. **b,** Computed bandstructure for the RPP n=4. **c,** Corresponding electron (dashed black lines) and hole (red) probability density profiles, and high frequency dielectric constant profile (blue) along the stacking axis ĉ. **d,** Calculated binding energy of the exciton ground state using the exciton reduced mass in Fig. 1e and the DFT results. (Gray dashed line) 3D perovskite value measured by Miyata et al.[16]. The binding energy of the excited exciton states (2s, 3s, 4s) were also computed and are reported in Supplementary Fig. 4.

Overall, theoretical solution to the exciton binding energy in each RPP requires the knowledge of: the electron and hole probability density distributions, the dielectric constant profile along the stacking direction, and the exciton effective mass. In RPPs, the density profiles and dielectric constants along the stacking direction (Fig. 2c and Supplementary Fig. 3b), along with the electronic bandstructure (Fig. 2b), were derived from the electron and hole wavefunctions calculated by DFT (Supplementary Text 1). A representative example of the electron and hole probability density profiles is sketched in Fig. 2c for the RPPs with n=4. Our calculations reveal that the charge wavefunctions stay confined into the perovskite layers and exhibit little leakage into the organic spacing layers, which is consistent with uncoupled quantum well systems. Moreover, the results reveal maximum values of dielectric constant in the range of 4 to 5.2 for the perovskite layers, and a minimum value of about 2.2 for the organic spacing layers, consistent with previous estimations[31,32]. The contrast of dielectric constants between the perovskite and organic spacers is at the origin of the strong dielectric confinement effects observed in RPPs, as discussed in the next section.

Gathering all these information, our general model was then applied to the calculation of the exciton binding energy in RPPs, yielding a value of ~435 meV for the RPP n=1, and which decreases monotonically to ~135 meV for n=4 as illustrated in Fig. 2d (see also Supplementary Fig. 4 and Supplementary Table 2). The accuracy of our model to predict the binding energies of excitons in 2D perovskites was then verified by directly measuring these values using optical spectroscopy techniques.

**Direct measurement of the exciton binding energy.** We directly measured the exciton binding energy in each RPP using optical absorption, photoluminescence (PL), and photoluminescence excitation (PLE) spectroscopy (Fig. 3). The optical bandgap was tuned over the visible spectral range from 2.540±0.004 eV (488 nm) in the RPP n=1 down to 1.846±0.004 eV (672 nm) for n=5 (Fig. 3a). We note that the RPPs with n>1 retained their optical bandgap from 290 K down to 4 K. This observation implies that our study conducted at low temperature, is directly transposable to room temperature and provides direct insights to the photo-physics of materials used in practical devices.

Fig. 3b shows a single sharp peak in the PL spectra, indicating the emission of the exciton ground state[14,33]. The low-energy shoulder observed more than tens of meV below the exciton peak in all RPPs, was recently assigned in the RPP n=1 to a phonon replica[34] and self-trapped excitons[35]. Based on power dependence measurements of the PL (Supplementary Fig. 5), we hypothesize a similar origin for the resonances in the RPPs with n>1, but we only focus on investigating the ground state exciton transition in this study.

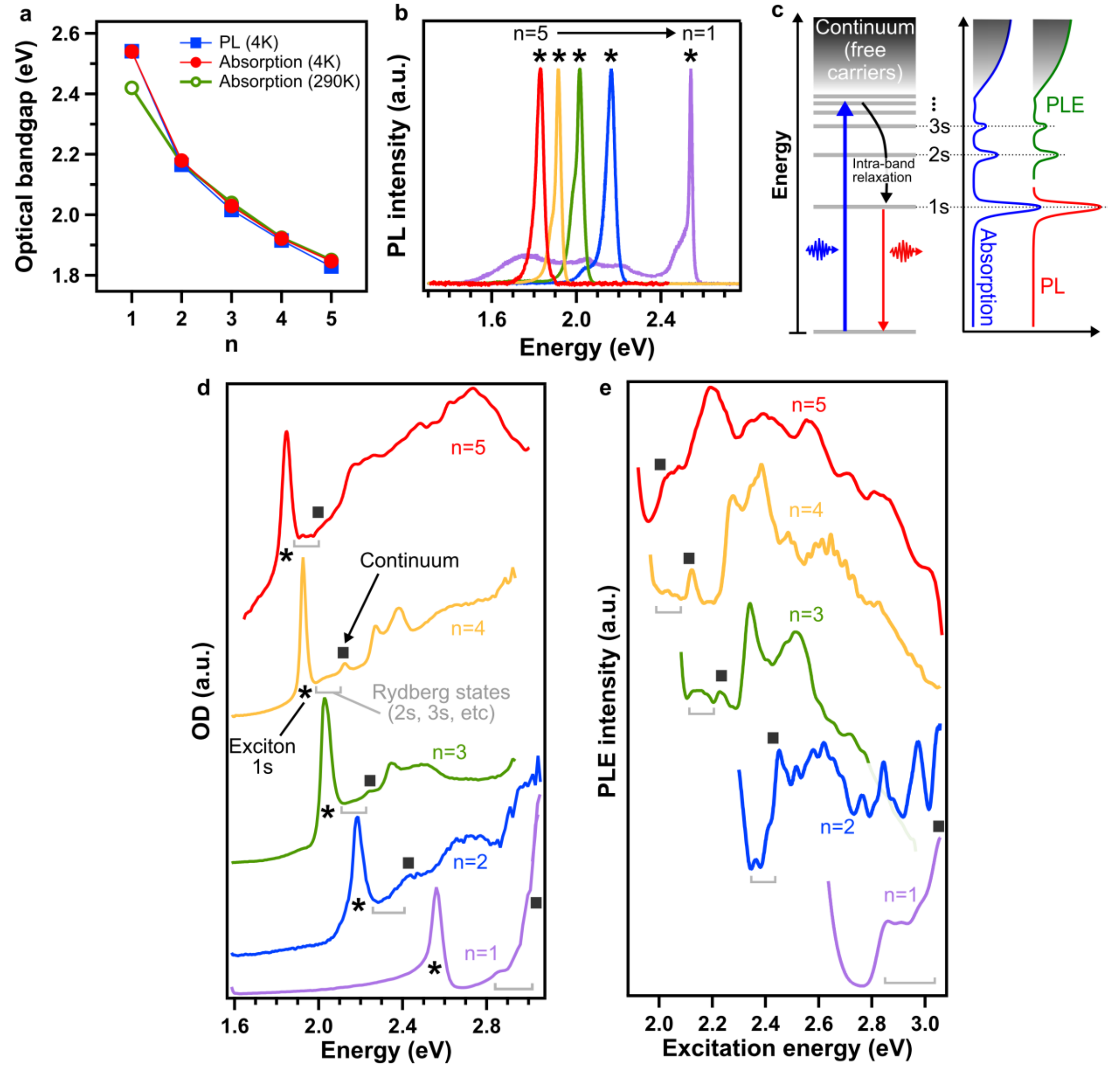

**Fig. 3 Optical spectroscopy of the RPP crystals with n=1 to 5. a,** Experimental optical bandgap scaling. **b,** Photoluminescence spectra**. c,** Schematics of the Rydberg series of the exciton ground state (1s) and excited exciton states (2s, 3s, etc.), and of the continuum. (right) Corresponding absorption or optical density, PL, PLE spectra typically observed in 2D material systems[26]. **d,** Optical density OD and **e,** PLE spectra of the RPPs with n=1 to 5, respectively. Stars point to exciton ground state optical transitions, black squares show the continuum onsets, and grey brackets indicate the region of excited exciton states.

All the presented absorption spectra show similar features (Fig. 3d), which includes a single absorption peak at low energy corresponding to the exciton ground state resonance. Moreover, a steady increase of absorption at higher energies modulated by energy optical transitions is well reproduced in the PLE measurements (Fig. 3e). These spectral features were assigned to the Rydberg series of the Wannier-Mott exciton (1s, 2s, 3s, 4s noted as Ns, with N=1, 2, 3, 4 and having energies $E_{Ns}$) and are illustrated in Fig. 3c. The onset of the continuum $E_G$ corresponds to electron and hole free carrier states (Fig. 3c), as previously reported in n=1 perovskites[15,33], other 2D nanomaterials[36], and quantum well semiconductors[26]. We note that absorption transition features observed in both the absorption and PLE spectra for energy higher than $E_G$ reflect the complex electronic bandstructure of RPPs[11,14,37], and which can be understood as transitions between valence and conduction bands away from the bandgap states (Fig. 2b and Supplementary Fig. 3a).

The exciton binding energies in RPPs were derived from the study of the optical transitions corresponding to the exciton Rydberg series and the continuum. We present a representative example for the RPP n=4 in Fig. 4a, b. Dielectric confinement (or image charge effect) has been shown[15,29,32,33] to mainly influence the 1s excitons in RPPs with n=1. Therefore, the 2s and 3s exciton states were fitted using the classic 2D hydrogen Rydberg series with energies[26] $E_{Ns}= E_G - R_y /(N-1/2)^2$, where $R_y$ is the Rydberg energy. This yields $E_G$=2.078±0.012 eV and $R_y$=0.11±0.04 eV for the RPP n=4 (Fig. 4b, dashed red line). The continuum energy is in good agreement with the apparent experimental value 2.100±0.010 eV. The same procedure was applied to the other RPPs (Supplementary Fig. 6). The binding energy of the 1s exciton ground state was derived from the difference $|E_{1s}-E_G|$, which ranges from about 470 meV for the RPP n=1 down to 125 meV for n=5 (Fig. 4c, d, Supplementary Table 1). We emphasise that the predicted theoretical and experimentally determined values are in excellent agreement (Fig. 4d and Supplementary Fig. 4), thus validating the accuracy of our developed theoretical model for Wannier-Mott excitons in RPPs. These results emphasize the need for an accurate determination of the exciton reduced mass and taking into account the contributions of both quantum and dielectric confinements in order to develop an accurate and robust theoretical model for the 2D perovskite systems.

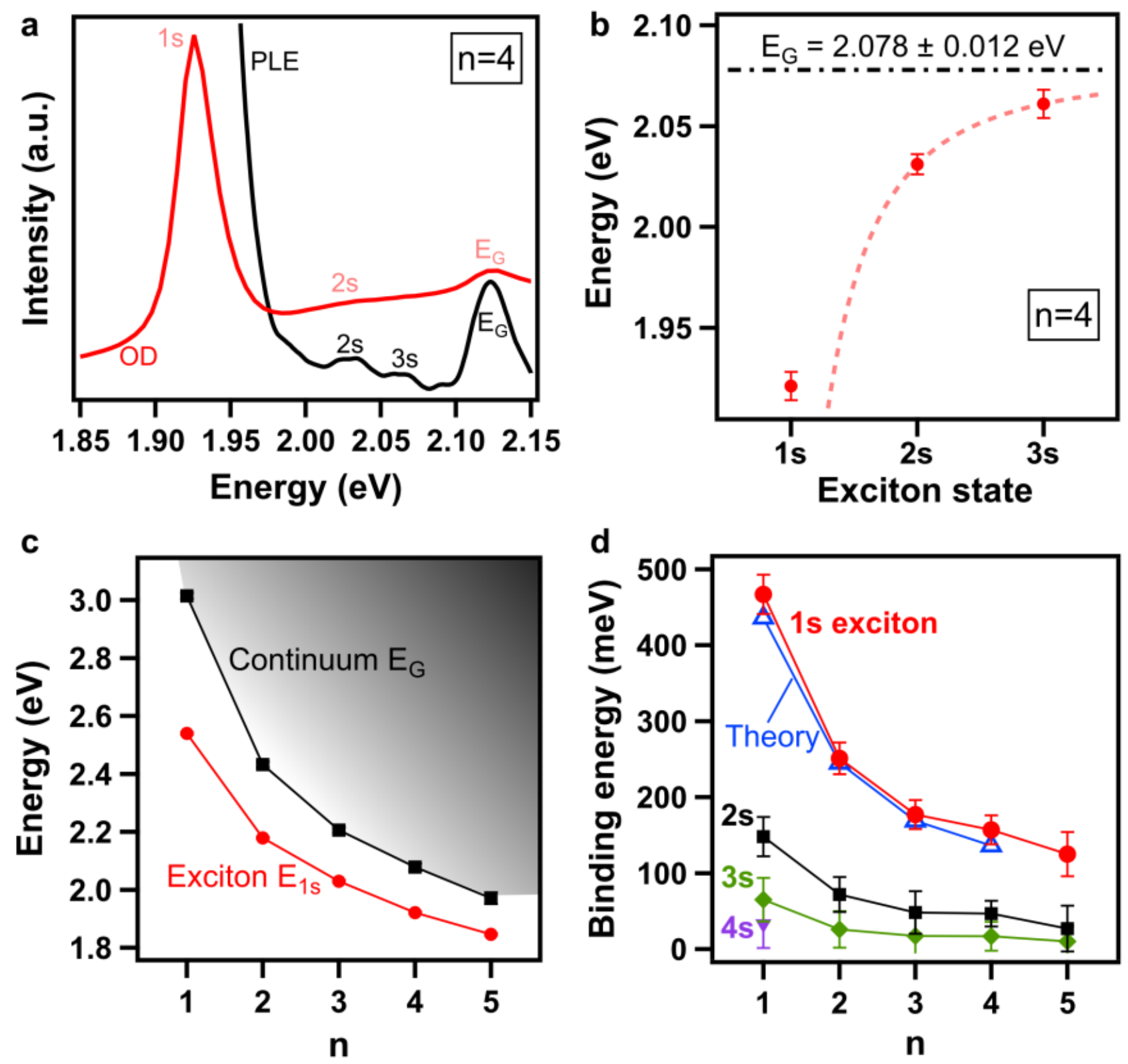

**Fig. 4 Direct measurement of the exciton binding energy and Rydberg states in RPPs with n=1 to 5. a,** Absorption OD and PLE of the RPP n=4 (see the others in Supplementary Fig. 6) clearly showing the exciton ground state 1s, the excited exciton states 2s and 3s, and the continuum onset $E_G$. **b,** Energy of the exciton Rydberg states for the RPP n=4. Dashed line is a fit to the 2s and 3s states with the 2D hydrogen model[26] of exciton Rydberg series using $R_y$=0.11±0.04 eV in the formula $E_{Ns}= E_G - R_y /(N-1/2)^2$, with Ns equals to 1s, 2s, 3s. **c,** Evolution of the exciton ground state and continuum energies with the 2D perovskite layer thickness (or n). **d,** Corresponding experimental binding energy of the exciton ground state (1s) and the excited exciton states (2s, 3s, 4s), and comparison to theoretical results for the exciton ground state.

After validating the model, we can further extract pertinent details on the specific role of dielectric confinement on the exciton characteristics and the scaling of the exciton binding energy with perovskite layer thickness. The model elucidates that the overlap of the electron and hole wavefunctions, whose convolution forms the exciton ground state, becomes more pronounced towards the centre of the perovskite layer for the highest n-value (Fig. 2c and Supplementary Fig. 3b). This suggests that the electric field lines for the exciton ground state are significantly more localized within the perovskite layer for larger n-values than for the RPP with n=1 and 2. This is consistent with the fact that with increasing perovskite layer thickness the strength of dielectric confinement decreases with respect to quantum confinement. Furthermore, the calculated probability densities of the electron and hole wavefunctions exhibit negligible intensity outside of the perovskite layer (i.e. into the organic spacer layer) as displayed in Fig. 2c and Supplementary Fig. 3b, which is comparable to a technologically relevant multi quantum well system. In summary, our theoretical model demonstrates that the strong exciton binding energy in thin 2D perovskite layers (n→1) stem from strong dielectric confinement effects, which wanes progressively for larger perovskite layer thicknesses (n→5 or above).

**Simple analytical expression of the exciton binding energy scaling law.** Finally, based on our understanding of the exciton behaviour with varying perovskite layer thickness, we developed a general scaling law of the exciton binding energies based on a classical model for low dimensional systems[38] as described in equation (1):

$$\mathrm{E}_{\mathrm{b,1s}} = \frac{\mathrm{E}_0}{\left(1+\frac{\alpha-3}{2}\right)^2} \text{ with } \alpha = 3 - \gamma e^{-\frac{L_W}{2\mathrm{a}_0}}. \quad (1)$$

In this expression for the exciton ground state binding energy, $E_0$ (=16 meV) and $a_0$ (=4.6 nm) are the 3D Rydberg energy and Bohr radius of 3D perovskites[16], respectively, and $L_w$ is the physical width of the quantum well (Supplementary Table 1) for an infinite quantum well potential barrier[39]. In the model (1), the exciton is considered isotropic in a α-dimensional space (1<α≤3) and we introduce a correction factor $\gamma$ in order to take into account dielectric confinement effects. More precisely, the factor γ depends on the dielectric constants of the perovskite quantum wells and organic spacer layers, and increases for larger dielectric contrast corresponding to stronger dielectric confinement effects. In a purely quantum confined regime $\gamma = 1$ in the expression of α, but experimentally derived α–values in RPPs can only be fitted with a larger $\gamma$ value ($\gamma = 1.76$) (Fig. 5a). The corresponding decrease in the value of α highlights additional compression of the exciton wavefunction in the quantum well due to dielectric confinement, which results in enhanced values of exciton binding energy as compared to merely including the quantum confinement effect (Fig. 5b). We note that in the simplistic approximation of Hydrogen exciton model, the changes of exciton confinement with perovskite layer thickness can be understood as changes of both the effective dielectric constant value and the exciton Bohr radius as a function of n value (Supplementary Fig. 7).

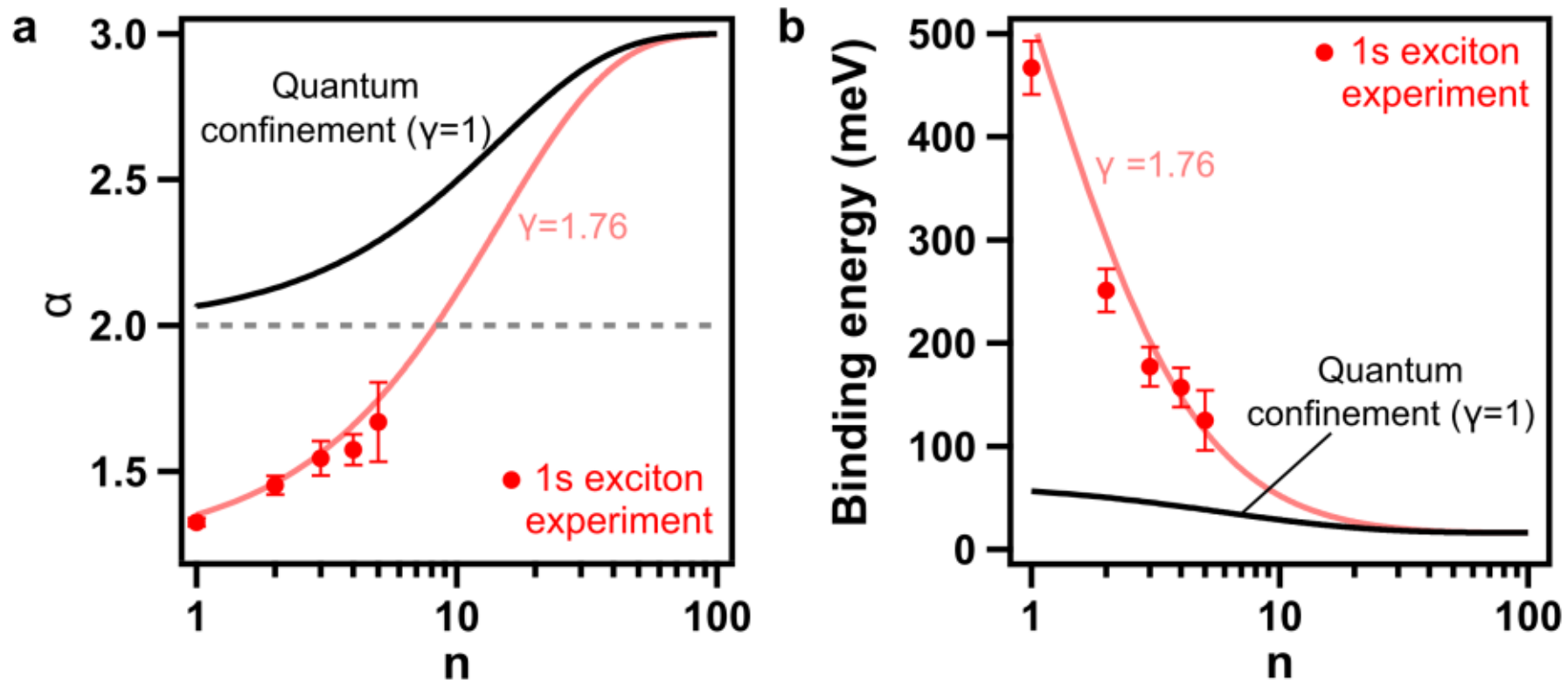

**Fig. 5 Scaling law of the exciton binding energy with the perovskite layer thickness. a,** the dimensionality coefficient α was derived from equation (6), where the exciton binding energy are the experimental values of Fig. 3d and $E_0$=16 meV, $a_0$=4.6 nm, and $L_w$= 0.6292×n in nanometers. (black curve) γ=1 corresponds to the case of pure quantum confinement in quantum well systems with infinite potential barriers. (red curve) γ=1.76 was derived from the fit to the experimental values of α (red markers) using the expression of α in equation (6). Setting γ>1 leads to a decrease of the value of α which reflects the more pronounced "compression" of the exciton ground state wavefunction in the perovskite layer due to dielectric confinement, as compared to the case of pure quantum confinement (γ=1). **b,** Corresponding results for the binding energy of the exciton ground state, showing the enhancement of the binding energy due to dielectric confinement. The red curve gives the general scaling law of the exciton binding energy with the perovskite layer thickness based on the equation (6), with $E_0$=16 meV, γ=1.76, $a_0$=4.6 nm, and $L_w$= 0.6292×n in nanometers.

Applying our model (1) for high n-values, we predict that the exciton binding energy in RPPs is larger than room temperature thermal fluctuations ($k_BT$) up to n~20 (perovskite slab thickness ~12.6 nm). This demonstrates the surprising robustness of exciton states in RPPs, inspite of the small spatial extend of the exciton as compared to the perovskite layer thickness. In other words, a more abrupt reduction of the exciton binding energy with increasing perovskite layer thickness was expected given the strong screening of the electron-hole Coulomb interaction reported in 3D perovskites[16], a situation approached by RPPs with perovskite layer thickness of a few nanometres. Again, this underline the unusual nature of photo-excited states in RPPs, and to more extend in solution-processed quantum well systems as a template for exploring quasi-2D semiconductors.

In summary, we demonstrate the importance of Coulomb interactions in 2D layered perovskites and experimentally elucidate properties of unexpectedly strongly bound excitons (>120 meV) in RPPs with thickness up to 3.1 nm (n=5). We propose a generic formulation of the scaling of the exciton binding energy with the perovskite layer thickness from the single layer (n=1) to 3D crystals ($n \rightarrow \infty$), further predicting the nature of optical transitions at room temperature to change from excitonic to free-carrier like only in RPPs with thickness larger than ~12 nm (n~20). These results mark a fundamental step towards the design of new 2D perovskite-based semiconductor materials for next generation optoelectronic and photonic technologies such as solar cells, light emitting diodes, photodetectors, polariton and electrically driven lasers.

## Methods

**RPP crystals synthesis and preparation.** The crystal structures of the RPPs, $(BA)_2(MA)_{n-1}Pb_nI_{3n+1}$, is composed of an anionic layer $\{(MA)_{n-1}PbnI_{3n+1}\}^{2-}$, derived from bulk methylammonium lead triiodide perovskites ($MAPbI_3$), which is sandwiched between n-butylammonium (BA) spacer cations (Fig. 1a). RPPs with n ranging from one to five, corresponding to perovskite layer thickness between 0.641 and 3.139 nm, were synthetized and purified following previously reported method[3–5]. More precisely, the raw crystals were prepared by combining PbO, MACl and BA in appropriate molar ratios in a $HI/H_3PO_2$ solvent mixture. The precursor solutions were prepared with 0.225 M of $Pb^{2+}$ concentration and stirred at room temperature overnight. Phase purity and crystalline quality of each crystal sample was established by monitoring x-ray diffraction (Supplementary Fig. 1). In addition, the small Stokes shifts and relatively sharp linewidths of the exciton resonances (Fig. 3) were another indication of the homogeneity and low disorder in the RPP crystals. Thin RPP crystals were mechanically exfoliated onto either the 3.5-μm-core of a single-mode optical fiber for magneto-absorption spectroscopy or transparent quartz substrates for optical absorption, PL, and PLE spectroscopy experiments.

**Magneto-absorption spectroscopy.** A single RPP crystal was affixed over the core of a 3.5 μm diameter single-mode optical fibre to ensure rigid optical alignment of the light path during the magnetic field pulse. The fibre-sample assembly was mounted in a custom optical probe that was fitted in the 4 K bore of a 65 T capacitor-driven pulsed magnet. The sample was in a helium exchange gas environment to ensure thermal anchoring at 4 K. White light from a Xe lamp transmits the sample via the single mode fibre and was retro-reflected and dispersed in a 300 mm spectrometer with a 300 groove/mm grating. Broadband spectra were recorded every 2.3 ms throughout the magnet pulse. Access to $\sigma^+$ and $\sigma^-$ circular polarization was achieved via a thin-film circular polarizer mounted directly after the sample and by reversing the direction of the magnetic field. Details of the setup can be found elsewhere[40].

**Optical absorption, PL, and PLE.** Optical spectroscopies were performed with an in-lab-built confocal microscopy system focusing close to the diffraction limit (~1-2 μm resolution) a monochromatic laser tunable over the visible and near-infra-red spectral ranges. PL spectral responses were obtained through a spectrograph (Spectra-Pro 2300i) and a CCD camera (EMCCD 1024B) yielding an error of less than 2 nm. PL data in the main text were measured for light excitation at 440 nm (if not mentioned otherwise) and the excitation intensity was typically of the order of or below $10^3$ mW/cm$^2$. PLE spectra were obtained by measuring the PL integrated intensity while scanning with a 2-nm-step the excitation light wavelength at wavelengths smaller than the one of the PL emission peak. Absorption spectra were measured either via a balanced photodiode by tuning the laser excitation wavelength or by detecting the spectral transmission/reflection of the samples exposed to white light. Samples were measured under vacuum ($10^{-5}$-$10^{-6}$ Torr) and cooled at 4 K if not mentioned otherwise.

**Theory.** A momentum space representation of the exciton Green's function developed previously for classic quantum wells[41] was used to solve the Bethe-Salpeter equation in the effective mass approximation. This method was adapted for RPPs to include the screening of the electron-hole interaction related to the dielectric confinement and the electron and hole wavefunctions overlap. These latter effects were evaluated at the DFT level. The Wannier-Mott exciton Rydberg states appear as the bound states in the absorption spectrum and can also be determined from the corresponding Schrödinger equation for the two particle wave functions (see details in Supplementary Text 1).

**Acknowledgments:** The work at Los Alamos National Laboratory (LANL) was supported by LDRD program (J-C.B, W.N, S.T, A.D.M) and was partially performed at the Center for Nonlinear Studies. The work was conducted, in part, at the Center for Integrated Nanotechnologies (CINT), a U.S. Department of Energy, Office of Science user facility. Part of this work was performed at the National High Magnetic Field Laboratory, which is supported by NSF DMR-1157490 and the State of Florida. Work at Northwestern University was supported by ONR grant N00014-17-1-2231. The work in France was supported by Agence Nationale pour la Recherche (TRANSHYPERO project). This work was granted access to the HPC resources of [TGCC/CINES/IDRIS] under the allocation 2017-A0010907682 made by GENCI.

**Author contributions:** J.-C.B and A.D.M conceived the idea, designed the experiments, and wrote the manuscript. J.-C.B made the samples, performed the optical spectroscopy measurements, analysed the data, and provided insights into the mechanisms with support from J.J.C. A.S performed the magneto-absorption measurements and analysed the data under the supervision of S.C. J.E developed the theoretical models. B.T performed the DFT calculations with support from C.K, L.P, and M.K. M.G.K, C.S.S developed the chemistry for the synthesis of phase-pure crystals with support from H.T and W.N. All authors contributed to this work, read the manuscript and agree to its contents, and all data are reported in the main text and supplemental information.